\documentclass[english]{emulateapj}
\usepackage[T1]{fontenc}
\usepackage[latin1]{inputenc}
\usepackage{graphicx}
\usepackage{amssymb}

\makeatletter

\usepackage{times}
\usepackage{amsmath}

\newcommand{\pers}{\,{\rm s^{-1}}}

\newcommand{\km}{{\,\rm km}}

\newcommand{\perGyr}{{\,\rm Gyr^{-1}}}
\newcommand{\pc}{\,\mathrm{pc}}
\newcommand{\Hz}{{\,\rm Hz}}

\newcommand{\Mbh}{M_{\bullet}}

\newcommand{\Mo}{M_{\odot}}

\newcommand{\Rs}{R_{\star}}
\newcommand{\Ms}{M_{\star}}
\newcommand{\ns}{n_{\star}}

\newcommand{\sigs}{\sigma_{\star}}

\newcommand{\LISA}{{\it LISA\,}}

\shorttitle{GRAVITATIONAL WAVE SOURCES}
\shortauthors{HOPMAN AND ALEXANDER}

\usepackage{babel}
\makeatother
\begin{document}
\bibliographystyle{apj.bst} 
%

%
%

\title{The effect of mass-segregation on gravitational wave sources
near massive black holes}

\author{Clovis Hopman and Tal Alexander\altaffilmark{1}}

\affiliation{Faculty of Physics, Weizmann Institute of Science, POB 26, Rehovot
76100, Israel} \altaffiltext{1}{The William Z. \& Eda Bess Novick career development chair}

\email{clovis.hopman, tal.alexander@weizmann.ac.il}

\begin{abstract}
Gravitational waves (GWs) from the inspiral of compact remnants (CRs)
into massive black holes (MBHs) will be observable to cosmological
distances. While a CR spirals in, 2-body scattering by field stars may
cause it to fall into the MBH before reaching a short period orbit
that would give an observable signal. As a result, only CRs very near
($\sim0.01\pc$) the MBH can spiral in successfully. In a multi-mass
stellar population, the heaviest objects sink to the center, where
they are more likely to slowly spiral into the MBH without being
swallowed prematurely. We study how mass-segregation modifies the
stellar distribution and the rate of GW events. We find that the
inspiral rate per galaxy for white dwarfs is $30\perGyr$, for neutron
stars $6\perGyr$, and for stellar black holes (SBHs) $250\perGyr$. The
high rate for SBHs is due to their extremely steep density profile,
$n_{\rm BH}(r)\propto r^{-2}$. The GW detection rate will be dominated
by SBHs.
\end{abstract}


\section{Introduction}

Massive Black Holes (MBHs) with masses $\Mbh\lesssim5\times10^6\Mo$
have Schwarzschild radii $r_S=2G\Mbh/c^2$, such that a test mass
orbiting at a few $r_S$ emits gravitational waves (GWs) with
frequencies $10^{-4}\Hz\!\lesssim\!\nu\!\lesssim\!1\Hz$, detectable by
the planned space based {\it Laser Interferometer Space
Antenna}\footnote{http://lisa.jpl.nasa.gov/} ({\it LISA}). Main
sequence (MS) stars with mass $\Ms$ and radius $\Rs$ will be disrupted
at the tidal radius $r_t=(\Mbh/\Ms)^{1/3}\Rs>r_S$ and are therefore
unlikely to be sources of observable GWs (our own Galactic center may
be an exception, Freitag \citeyear{Fre03}). Compact remnants (CRs)
such as white dwarfs (WDs), neutron stars (NSs) and stellar black
holes (SBHs) have tidal radii $r_t<r_S$ and can emit GWs that are
observable to cosmological distances. The inspiral of a CR into a MBH
(``extreme mass ratio inspiral sources'' [EMRIs]) is among the main
targets of \LISA.

The event rate of EMRIs has been estimated by numerous authors
\citep{Hil95, Sig97b, Mir00, Fre01, Iva02, Fre03, Ale03b, Hop05,
Hop06} but remains rather uncertain, in part because of the slow
nature of the inspiral process, which occurs on many dynamical
times. This makes the inspiraling star very susceptible to scattering
by other stars, which can change the orbital parameters. The
formalism for inspiral rates is similar to that for prompt consumption
of stars \citep{Bah77, Lig77, Fra76, Coh78, Sye99, Mag99}, but there
are some important differences because the process is much slower.

The picture can be understood as follows: Let $t_r$ be the relaxation
time of a star with negative energy $E$ (hereafter ``energy''; $E>0$
for bound stars) and specific angular momentum $J$ (hereafter
``angular momentum''). The relaxation time is the time-scale for a
change of energy of order $E$, or a change in angular momentum of
order $J_c(E)$, the circular angular momentum. The change in $J$ of a
star per orbital period $P$ is $\Delta J = (P/t_r)^{1/2}J_c$. The
time-scale for a change of order $J$ is $t_J=(J/J_c)^2t_r$. In
particular, the time-scale for a change in $J$ by the order of the
loss-cone, determined by the angular momentum of the last stable orbit
$J_{\rm LSO}=4G\Mbh/c$, is $t_{lc}=(J_{\rm LSO}/J_c)^2t_r$. Inspiral
due to dissipation by GW emission happens on a time-scale $t_0(E,J)$,
which for highly eccentric orbits has a very strong angular momentum
dependence, $t_0(J)\propto J^{7}$. If $t_{lc}\ll t_0(E, J\!\to\!
J_{\rm LSO})$, the angular momentum will be modified even if the star
has $J\gtrsim J_{\rm LSO}$. As a result it is very likely that the
star will be scattered into the loss-cone (or away from it, to an
orbit where energy dissipation is very weak). Such CRs will eventually
be consumed by the MBH and add to its mass, but they will not be
observable as GW emitters (GW bursts in our own GC may form an
exception [Rubbo, Holley-Bockelmann \& Finn \citeyear{Rub06}]).

The approximate condition $t_0(E, J\!\to\! J_{\rm LSO})\!<\!t_{lc}(E)$
translates into a minimal energy or maximal semi-major axis $a_{\rm
GW}$ a CR must have in order to spiral in and become a \LISA source
(``successful inspiral''); Hopman \& Alexander (\citeyear{Hop05};
hereafter HA05) estimate that for a MBH of $\Mbh\!=\!3\times10^6\Mo$,
$a_{\rm GW}\!\sim\!0.01\pc$: nearly all CRs with $a\!\gg\! a_{\rm GW}$
are promptly captured or deflected without giving an observable
signal, while nearly all stars with $a\!\ll\! a_{\rm GW}$ do spiral in
successfully.

The fact that the distribution of CRs near MBHs is crucial to the
observational outcome, implies that mass-segregation is likely to play
a very important role for EMRIs. Mass-segregation is a manifestation
of dynamical friction. It drives the heaviest objects to the center,
so their concentration within $a_{\rm GW}$ increases, and drives the
lightest stars to larger radii, so that they are relatively rare
within $a_{\rm GW}$. The importance of mass-segregation on inspiral
processes was dramatically demonstrated in $N$-body simulations
\citep{Bau04b, Bau05b} of tidal capture of MS stars \citep{Ale03a,
Hop04}. \citet{Bau05b} studied tidal capture by a $\sim10^3\Mo$ black
hole in a young stellar cluster with MS masses up to $\sim100\Mo$. In
spite of the fact that massive stars are scarce, captured stars
typically had masses $\Ms\sim20\Mo$.

In this Letter we study the implications of mass-segregation on the
EMRI rate.

\section{Model}\label{s:model}

Our model is based on \citet{Bah76, Bah77}. Here we briefly
recapitulate the main assumptions, and discuss our treatment of GW
capture. A detailed discussion of our model can be found in HA05 and
Hopman \& Alexander (\citeyear{Hop06}).

\subsection{Dynamics}

The MBH dominates the dynamics of stars within its {}``Bondi radius'',
or radius of influence, $r_{h} = {G\Mbh/\sigs^{2}}$, where $\sigs$ is
the velocity dispersion of a typical star of mass $\Ms\ll\Mbh$
(assumed of Solar type), which we will use to scale our
expressions. Orbits are assumed to be Keplerian within $r_h$. Each
species with mass $M$ is described by a distribution function (DF) in
energy space $f_{M}(E)$.

We define a dimensionless time $\tau=t/T_{h}$ in terms of the relaxation
time at the radius of influence
\begin{equation}
T_{h} = {3(2\pi\beta/\Ms)^{3/2}\over 32\pi^2 G^2 \Ms^2 \ns
\ln\Lambda},
\end{equation}
where $\ns$ is the number density at $r_{h}$ for the typical star
$\Ms$, $\beta\!=\!\Ms\sigs^2$, and $\Lambda\!=\!\Mbh/\Ms$. Introducing
the dimensionless energy $x\!=\!(\Ms/ M)(E/\beta)$ and the
dimensionless DF
$g_{M}(x)\!=\![(2\pi\beta/\Ms)^{3/2}\ns^{-1}]f_{M}(E)$, the
Fokker-Planck equation in energy space is (Bahcall \& Wolf
\citeyear{Bah77} Eq. [26])

\begin{equation}\label{e:dgmdt}
{\partial g_M(x, \tau)\over\partial\tau} =  -x^{5/2} {\partial \over \partial x}Q_{M}(x) - R_{M}(x).
\end{equation}
We also write equation (\ref{e:dgmdt}) in a logarithmic form suitable
for numerical integration (see appendix). The spatial number density
$n_M(r)$ of stars is related to the DF by

\begin{equation}
n_M(r) = 2\pi^{-1/2}\ns\int_{-\infty}^{r_h/r}dx g_{M}(x)\left[r_h/r-x\right]^{1/2}.
\end{equation}
We fit our numerical results by power-laws $n_M(r)\propto r^{-\alpha_{M}}$.

In expression (\ref{e:dgmdt}), $Q_{M}(x)$ is the (dimensionless)
rate at which stars flow to energies larger than $x$,

\begin{eqnarray}\label{e:Qm}
Q_{M}(x) &=&  \sum_{M'}{M\over \Ms} {M'\over \Ms}  \int_{-\infty}^{x_{D}}dx'\left[\max\left(x,x'\right)\right]^{-3/2}\times\nonumber\\
&&\times\left[ g_M(x){\partial g_{M'}(x')\over\partial x'} - {M'\over M}g_{M'}(x'){\partial g_M(x)\over\partial x}\right].
\end{eqnarray}

The dimensional stellar current is related to $Q_{M}$ by
$I_{M}(E,t)=I_{0}Q_{M}(x,t)$, where
\begin{equation}
I_{0}\equiv{\frac{8\pi^{2}}{3\sqrt{2}}}r_{h}^{3}\ns{\frac{(G\Ms)^{2}\ln\Lambda \ns}{\sigs^{3}}}\,.\label{e:I0}\end{equation}
\citep{Bah76, Hop06}.

The last term in equation (\ref{e:dgmdt}) represents losses of stars
due to loss-cone effects (both prompt infall and inspiral) in
$J$-space. The sink term in the diffusive regime for the loss-cone
is

\begin{equation}\label{e:Rm}
R_{M}(x) = {g_M(x)\over \tau_r(x)\ln[J_c(x)/J_{\rm LSO}]},
\end{equation}
where $J_c(x)/J_{\rm LSO}=(1/4\sqrt{2})(c/\sigs)x^{-1/2}$. The
full-loss cone regime, $x\lesssim10$, does not contribute to the GW
event rate (Alexander \& Hopman \citeyear{Ale03b}; HA05; Hopman \&
Alexander \citeyear{Hop06}). In our calculations we neglect the sink
term in the full loss-cone regime by setting $R_{M}\to0$ for
$x<10$. In Eq. (\ref{e:Rm}) the dimensionless {\it local relaxation
time} $\tau_r(x)$ is
\begin{equation}
\tau_r(x) = {\Ms^{2}\over\sum_{M}g_M(x)M^{2}},
\end{equation}
independent of the stellar mass.

Let the number of stars accreted to the MBH before giving an
observable GW signal be $N_p(x)$, and the number of those that spiral
in successfully and do give a signal $N_i(x)$. The steady state result
for $\tau_r(x)$ is used to determine the probability for inspiral
$S_M(x)=N_i(x)/[N_i(x) + N_p(x)]$ by Monte Carlo simulations (HA05) as
follows. At every orbit, a star of initial energy $E$ and initially
large $J$ makes a step in $J$ of order $\Delta J =
[P(x)/t_r(x)]^{1/2}J_c$ with random sign because of scattering, and
loses energy $\Delta E_{\mathrm{GW}}=(85\pi/24576)(M/\Mbh)M
c^2(J/J_{\rm LSO})^{-7}$ to GWs \citep{Pet64}. This is repeated many
times and the outcome is recorded. The total rate of {\it successful
inspirals} for species $M$ is then given by $\Gamma_{M} =
I_0\int^{\infty} dxS_M(x)x^{-5/2}R_{M}(x)$. It is convenient to
express the capture rate in terms of the semi-major axis $a=r_h/2x$ of
the stars,

\begin{equation}\label{e:Gam}
\Gamma_{M}(<a) = {2\sqrt{2}I_0\over r_h^{3/2}}\int_{0}^{a}daa^{1/2}S_M(a)R_{M}\left(a\right).
\end{equation}

\subsection{Boundary conditions and model parameters}

Equation (\ref{e:dgmdt}) has inner and outer boundary conditions. At
some large energy $x_{D}$ the DF vanishes, $g(x\!>\!x_D)\!=\!0$. Since
the EMRI rate is dominated by the largest distance where successful
capture is possible, the exactly value of $x_D$ is not important. Here
we used $x_D=10^4$, which is approximately the energy-scale where the
inspiral time becomes smaller than a Hubble time even for a circular
orbit. We assume that the MBH mass is $\Mbh=3\times10^6\Mo$,
representative of a typical \LISA source. Our value of $x_D$ would
approximately correspond to a distance scale of $\sim 10^{-4}\pc$ from
the MBH.

The second boundary condition is given at $x=0$: following
\citet{Bah77}, we assume that for $x<0$ the stars have a Maxwellian
velocity DF with equal temperature ($\beta_{M}\equiv
M\sigma_{M}^2=\beta$), and with different population number fractions
for different species $C_{M}$,

\begin{equation}
g_{M}(x) = C_M\exp(Mx/\Ms);\,\,\,\,\,\,\,\,\,\,\,(x<0).
\end{equation}
We consider four populations of stars. One species consists of main
sequence stars, assumed here to be of Solar mass. MSs do not
contribute to the GW inspiral rate since they are tidally disrupted
before spiraling in, but they do contribute dynamically and they
dominate both in number and in total mass at the radius of
influence. The other three populations consists of WDs ($M_{\rm
WD}=0.6\Mo$), NSs ($M_{\rm NS}=1.4\Mo$) and SBHs ($M_{\rm
BH}=10\Mo$). The number fraction ratios of the four populations at
$x=0$ are $C_{\rm MS}:C_{\rm WD}:C_{\rm NS}:C_{\rm
BH}=1:0.1:0.01:10^{-3}$, typical for continuously star forming
populations \citep{Ale05}. We also adopt for our model the Galactic
center values $\sigs=75\km\pers$ ($r_h=2\pc$) and
$\ns=4\times10^4\pc^{-3}$ \citep{Gen03a}\footnote{The Galactic MBH
obeys the $\Mbh-\sigma$ relation \citep{Fer00, Geb00, Tre02}, so that
these values may also be representative of other MBHs.}. The model
parameters are summarized in table (\ref{t:model}).

\section{Results}\label{s:results}

\begin{figure}
\plotone{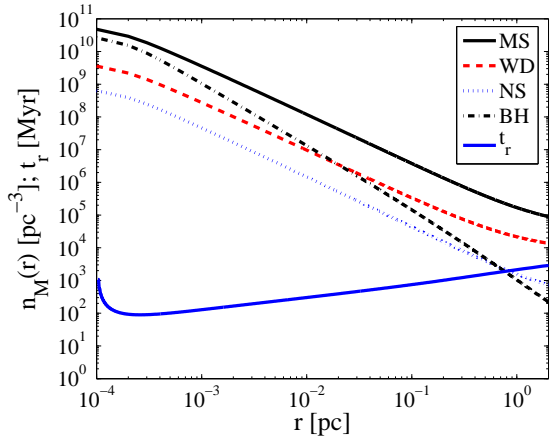} \figcaption[FileName]{Steady state number
densities $n_{M}(r)$. The DF of SBHs (dashed-dotted) is much steeper
than that of the other three stars ($\alpha_{\rm
BH}\!\approx\!2$). SBHs dominate NSs (dotted) by number nearly
everywhere, and they even dominate WDs (red dashed) within
$\sim0.01\pc$, in spite of their low number fraction at $r_h$, $C_{\rm
BH}/C_{\rm WD}=1\%$. MS stars (solid) dominate by number
everywhere. For the other three species we found $\{\alpha_{\rm MS},
\alpha_{\rm WD}, \alpha_{\rm NS}\}\!=\!\{1.4,1.4,1.5\}$. The
relaxation time (rising solid line) grows approximately as
$t_r\!\propto\!  r^{0.5}$.\label{f:nr}}
\end{figure}

We integrated Eq. (\ref{e:dgmdt}) until steady state is obtained,
after time $\tau\lesssim1$. In figure (\ref{f:nr}) we show the
resulting densities for the different species.  The DF of the SHBs is
much steeper than that of the other types ($\alpha_{\rm BH}=2.0$), and
at $r\approx0.01\pc$ the number density of SBHs becomes comparable to
that of the WDs. MS stars dominate everywhere by number, although we
did not take into account stellar collisions \citep{Fre02, Fre05}
which could deplete the MSs close to the MBH. SBHs also determine the
functional behavior of $t_r\propto r^{p}$, where
$p\!\approx\!\alpha_{\rm BH}\!-\!3/2\!\approx\!0.5$.

Throughout most of the cusp $\alpha_{\rm BH}\!\gtrsim\!2$, but near
$x_D$ the DF flattens, as required for the integrals in equation
(\ref{e:Qm}) to converge at high energies \citep{Bah77}. Large slopes
at intermediate energies are allowed by these equations, and arise
when a population of massive objects with a low number density exists,
as is the case in our model. At low energies the massive objects sink
effectively to the center by dynamical friction. At high energies the
massive objects dominate the dynamics, decouple from the lighter
objects, and form an $\alpha=7/4$ ``mini-cusp''. This process is
reminiscent of the Spitzer instability in globular clusters, where
SBHs decouple from the other stars \citep{Spi71, Kha06}.

\begin{figure}
\plotone{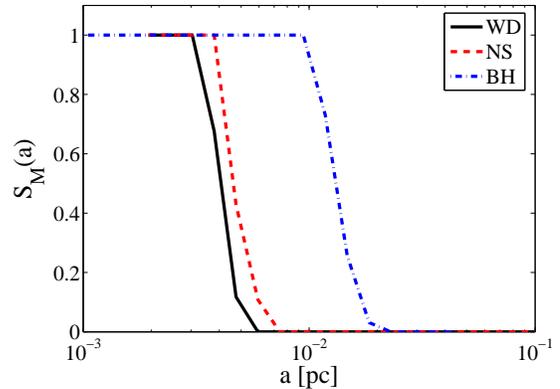} \figcaption{Probability of successful inspiral for a
consumed star as a function of distance from the MBH for the CRs. BHs
(dotted-dashed) can successfully spiral in from further distances
than WDs (solid) and NSs (dashed) due to their higher
masses. \label{f:Sa}}
\end{figure}

The probability for inspiral $S_{M}(a)$ is shown in figure
(\ref{f:Sa}). Since the SBHs are more massive they lose energy to GWs
at a higher rate than the other species and can spiral in from larger
distances.

\begin{figure}[!h]
\plotone{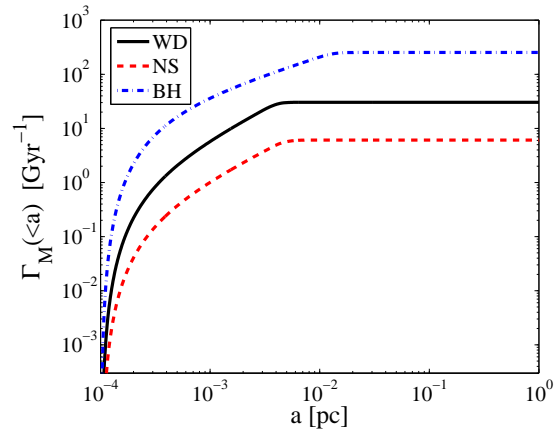}
\caption{Cumulative rates as a function of $a$ are dominated by BHs
(dotted-dashed); WDs (solid) and NSs (dashed) contribute significantly
less.\label{f:tr}}
\end{figure}

In figure (\ref{f:tr}) we show the cumulative rates of successful
inspiral (Eq. \ref{e:Gam}) for all CRs as a function of distance from
the MBH. We summarize some results in table (\ref{t:model}), where we
also give the enclosed number of stars $N_{M}(<a)$ within $a$.

\begin{table}[t]
\caption{Model parameters and GW rates}\label{t:model}
\begin{tabular}{llllllll}
\hline
\hline
Star   &   $M$    & $C_M$      & $\alpha_{M}$ & $N_{M}(\!<\!0.01\!\pc)$& $N_{M}(\!<\!0.1\!\pc)$  
  & $a_{\rm GW}$ & $\Gamma_{M}$\tabularnewline
       &$[\Mo]$   &            &          &                  &                     &  [mpc]       & $[{\rm Gyr^{-1}}]$\tabularnewline
\hline
MS     &  1       & 1.0        &    1.4   &    $10^3$        &  $3\times10^4$      & -            & -   \tabularnewline
WD     &  0.6     & 0.1        &    1.4   &     80           &  $2.7\times10^3$    & 4            & 30  \tabularnewline
NS     &  1.4     & 0.01       &    1.5   &     12           &  374                & 5            & 6   \tabularnewline
SBH    &  10      & $10^{-3}$  &    2.0   &    150           &  $1.8\times10^3$    & 13           & 250 \tabularnewline
\hline
\label{t:t1}
\end{tabular}
\end{table}

\section{Summary and discussion}\label{s:disc}

In our model, SBHs dominate the EMRI rate, in spite of their small
number density at $r_h$. The combination of a very steep cusp
($\alpha_{\rm BH}\approx2.0$) and a larger $a_{\rm GW}$ due
to their larger mass leads to $\Gamma_{\rm BH}>(\Gamma_{\rm WD},
\Gamma_{\rm NS})$ per galaxy. We also note that the amplitude of the GWs
is proportional to the stellar mass, so that the distance at which
these objects can be observed is $\!\sim\!10$ times larger than that
for WDs and NSs. Thus, SBHs will dominate the cosmic detection rate.

It is instructive to compare the EMRI rates we obtain here to those
obtained by HA05, where mass-segregation was not explicitly
included. For SBHs, $\Gamma_{\rm BH}$ is larger by a factor
$\sim50$. Part of the difference is that we assume here a larger total
number of SBHs within the cusp (by a factor $\sim6$; we normalized the
SBH number fraction at $r_h$ to be $C_{\rm BH}=10^{-3}$, while HA05
assumed that the {\it enclosed} fraction of SHBs is $10^{-3}$). More
importantly, the steeper cusp leads to a higher capture rate (by a
factor $\sim9$, HA05, eq. [32]; HA05 assumed $\alpha_{\rm
BH}=1.75$). The BH cusp is much steeper than any of the cases studied
by \citet{Bah77}, and in particular it is steeper than the cusp of a
single mass population, $\alpha=7/4$ \citep{Bah76}.

The rates $\Gamma_{\rm WD}$ and $\Gamma_{\rm NS}$ are also somewhat
larger than those found by HA05. Here the difference originates mainly
in the behavior of $t_r$: For WDs and NSs, $t_r$ was assumed to be
constant by HA05, as appropriate for a single mass population with
$\alpha=3/2$. However, the interaction between SBHs and the other CRs
leads to a decrease in $t_r$ towards the MBH (Fig. \ref{f:nr}). Using
the analytical expressions by HA05, it can be shown that if
$n_{M}(r)\propto r^{-3/2}$, and $t_r\propto r^{p}$, the successful
inspiral rate is enhanced by $(d_c/r_h)^{-3p/(3-2p)}\sim10$ (for
$p\!=\!0.5$) relative to the $t_r={\rm const.}$ case, where
$d_c=[(85/3072)\sqrt{G\Mbh}(M/\Mbh)t_h]^{2/3}$, see HA05 eq. (29).

The EMRI rates we found here are promising for the \LISA detection
rate \citep{Bar04a, Gai04}, in spite of the fact that more sources
also imply a stronger background noise \citep{Bar04b}.

We neglected here the effect of resonant relaxation \citep{Rau96,
Rau98}, which can increase the EMRI rate by up to an order of
magnitude \citep{Hop06}. A multi-mass analysis of RR has yet to be
performed. In addition to {\it direct capture} of CRs, EMRI can occur
following the formation of SBHs in accretion disks \citep{Lev03a},
binary disruptions \citep{Mil05} and tidal capture followed by a super
nova explosion of the captured star \citep{Hop05b}. These other
mechanisms lead to low eccentricity signals, whereas direct capture
leads to high eccentricities (HA05).

Our estimate of the number fraction of unbound SBHs is somewhat
uncertain, in part because we neglected dynamical effects for unbound
stars. We note that our estimate $N_{\rm
BH}(<\!\!\pc)\sim1.6\times10^4$ is consistent with calculations by
\citet{Mir00}, who found $N_{\rm BH}(<\!\!\pc)\sim2.5\times10^4$. Our
Galactic Center contains a MBH of mass comparable to the MBH mass
considered here \citep{Ghe05, Eis05}. Observational effects of a
cluster of SBHs near the Galactic MBH include microlensing
\citep{Cha01}, X-ray emission \citep{Pes03}, capture of massive stars
by an exchange interaction \citep{Ale04} and deviations from Keplerian
motion of luminous stars \citep{Mou05}. Such effects could in
principle be used to constrain the predicted densities.

\acknowledgements{TA is supported by ISF grant 295/02-1, Minerva grant
8563, and a New Faculty grant by Sir H. Djangoly, CBE, of London, UK.}

\appendix{}

Because of the large range of energies, the natural way to integrate
the Fokker-Planck equation is to divide the energy range into equal
logarithmic intervals. For convenience we give here the equations in
terms of the logarithmic distance variable $z=\ln(1+x/\lambda)$. The
Fokker-Planck equation without sink terms is then written as

\begin{eqnarray}\label{eq:Qi}
{\frac{\partial g_M(z,\tau)}{\partial\tau}}&=&-{M\over \Ms}(e^{z}-1)^{5/2}e^{-z}\times\nonumber\\
&&\sum_{M'} {M'\over \Ms}{\frac{\partial}{\partial z}}\Big[(e^{z}-1)^{-3/2}g_M(z)[g_{M'}(z)-g_{M'}(-\infty)] + g_M(z)\int_{z}^{z_D}dz'(e^{z'}-1)^{-3/2}{\partial g_{M'}(z')\over \partial z'}\nonumber\\
&& -  {M'\over M}(e^{z}-1)^{-3/2}e^{-z}{\partial g_M(z)\over \partial z}\int_{-\infty}^{z}dz'e^{z'}g_{M'}(z')  -  {M'\over M}e^{-z}{\partial g_M(z)\over \partial z}\int_{z}^{z_D}dz'e^{z'}(e^{z'}-1)^{-3/2}g_{M'}(z').\Big]
\end{eqnarray}

The logarithmic expressions for the sink term (eq. \ref{e:Rm}) can be
included directly by replacing $x\rightarrow \lambda(e^{z}-1)$.


\end{document}